\documentclass[twocolumn,showpacs,aps,prb,superscriptaddress,floatfix]{revtex4-1}
\usepackage{amsmath,amssymb}
\usepackage{latexsym}
\usepackage{graphicx}% Include figure files
\usepackage{color}
\usepackage{fancyhdr}
\usepackage{gensymb}
\usepackage{dcolumn}% Align table columns on decimal point
\usepackage{natbib}

\begin{document}

\title{Quenched crystal field disorder and magnetic liquid ground states in Tb$_2$Sn$_{2-x}$Ti$_x$O$_7$ }

\author{B.D.~Gaulin}
\affiliation{Department of Physics and Astronomy, McMaster University, Hamilton, Ontario, L8S 4M1, Canada}
\affiliation{Canadian Institute for Advanced Research, 180 Dundas St. W., Toronto, Ontario, M5G 1Z8, Canada}
\affiliation{Brockhouse Institute for Materials Research, McMaster University, Hamilton, Ontario, L8S 4M1, Canada}

\author{E.~Kermarrec}
\affiliation{Department of Physics and Astronomy, McMaster University, Hamilton, Ontario, L8S 4M1, Canada}
\affiliation{Laboratoire de Physique des Solides, Bat.510, Universit\'{e} Paris-Sud 11, UMR, CNRS 8502, F-91405, Orsay, France}

\author{M.L.~Dahlberg}
\affiliation{Department of Physics and Materials Research Institute, Pennsylvania State University, University Park, PA, 16803, USA}
\affiliation{The National Academies, 500 Fifth Street NW, Washington, DC, 20001, USA}

\author{M.J.~Matthews}
\affiliation{Department of Physics and Materials Research Institute, Pennsylvania State University, University Park, PA, 16803, USA}

\author{F.~Bert}
\affiliation{Laboratoire de Physique des Solides, Bat.510, Universit\'{e} Paris-Sud 11, UMR, CNRS 8502, F-91405, Orsay, France}

\author{J.~Zhang}
\affiliation{Department of Physics and Astronomy, McMaster University, Hamilton, Ontario, L8S 4M1, Canada}

\author{P.~Mendels}
\affiliation{Laboratoire de Physique des Solides, Bat.510, Universit\'{e} Paris-Sud 11, UMR, CNRS 8502, F-91405, Orsay, France}
\affiliation{Institut Universitaire de France, 103 bd Saint Michel, F-75005 Paris, France}

\author{K.~Fritsch}
\affiliation{Department of Physics and Astronomy, McMaster University, Hamilton, Ontario, L8S 4M1, Canada}
\affiliation{Helmholtz-Zentrum Berlin für Materialien und Energie, Hahn-Meitner-Platz 1, 14109 Berlin, Germany}

\author{G.E.~Granroth}
\affiliation{Quantum Condensed Matter Division, Oak Ridge National Laboratory, Oak Ridge, TN, 37831, USA}

\author{P.~Jiramongkolchai}
\affiliation{Department of Chemistry, Princeton University, Princeton, NJ, 08544, USA}

\author{A.~Amato}
\affiliation{Laboratory for Muon Spin Spectroscopy, Paul Scherrer Institute, CH-5232 Villigen PSI, Switzerland}

\author{C.~Baines}
\affiliation{Laboratory for Muon Spin Spectroscopy, Paul Scherrer Institute, CH-5232 Villigen PSI, Switzerland}

\author{R.J.~Cava}
\affiliation{Department of Chemistry, Princeton University, Princeton, NJ, 08544, USA}

\author{P.~Schiffer}
\affiliation{Department of Physics and Materials Research Institute, Pennsylvania State University, University Park, PA, 16803, USA}
\affiliation{Department of Physics, University of Illinois, Urbana, IL, 61801-3080, USA}

\begin{abstract}

Solid-solutions of the ``soft" quantum spin ice pyrochlore magnets Tb$_2$B$_2$O$_7$ with B=Ti and Sn display a novel magnetic ground state in the presence of strong B-site disorder, characterized by a low susceptibility and strong spin fluctuations to temperatures below 0.1 K. These materials have been studied using ac-susceptibility and $\mu$SR techniques to very low temperatures, and time-of-flight inelastic neutron scattering techniques to 1.5 K. Remarkably, neutron spectroscopy of the Tb$^{3+}$ crystal field levels appropriate to at high B-site mixing ($0.5<x<1.5$ in Tb$_2$Sn$_{2-x}$Ti$_x$O$_7$) reveal that the doublet ground and first excited states present as continua in energy, while transitions to singlet excited states at higher energies simply interpolate between those of the end members of the solid solution. The resulting ground state suggests an extreme version of a random-anisotropy magnet, with many local moments and anisotropies, depending on the precise local configuration of the six B sites neighboring each magnetic Tb$^{3+}$ ion.

\end{abstract}

\maketitle

\section{Introduction}

The rare earth pyrochlores have been a focus for the study of the physics of geometrical frustration, 
as their magnetic moments decorate a network of corner-sharing tetrahedra \cite{gardner2010magnetic}, one of the canonical building blocks for frustration in three dimensions \cite{review}. In combination with appropriate magnetic couplings and anisotropies, exotic disordered ground states often result.  In particular Ho$_2$Ti$_2$O$_7$ \cite{harrisbramwell,fennell2009magnetic,clancy2009revisiting} and Dy$_2$Ti$_2$O$_7$ \cite{Ramirez} exhibit net ferromagnetic coupling and strong local Ising anisotropy, resulting in a disordered spin ice ground state \cite{bramwell2001spin, balents2010spin} with Coulombic spin correlations and elementary excitations that behave as magnetic monopoles \cite{castelnovo2008magnetic,jaubert2009signature, fennell2009magnetic, tennant}.   Strong quantum effects are not manifestly displayed by these ``classical" spin ice materials, and interest has now focused on several candidate pyrochlore magnets where a quantum analogue of spin ice physics may be manifest.  Yb$_2$Ti$_2$O$_7$, for example, possesses a spin Hamiltonian with a leading Ising spin component, as in classical spin ice, but with an effective $S=1/2$ quantum moments arising from an XY-like g-tensor, and substantial exchange couplings that induce quantum dynamics \cite{RossX}. 

In (Ho, Dy, Yb)$_2$Ti$_2$O$_7$, the ground state crystal field (CF) doublet, which describes the rare earth moment in its pyrochlore environment (the Ti$^{4+}$ is non-magnetic), is well separated from the next-highest 
energy CF state, rendering all but the ground state doublet irrelevant to their low temperature physics.  However, this is not the case for Tb$^{3+}$ in Tb$_2$Ti$_2$O$_7$, wherein the $J=6$ total angular momentum multiplet splits into $(2J+1)=13$ states, the lowest energy of which are a ground state doublet and an excited state doublet roughly 1.2 meV above it \cite{ZhangCEF, GingrasCEF, Mirebeau2007}.  Tb$_2$Ti$_2$O$_7$ displays antiferromagnetic interactions with a Curie-Weiss constant of $\sim$ -19 K \cite{Gardner99, GardnerPRB, Mirebeau2007}, but no conventional long range order to below $\sim$ 0.1 K, even though it would be expected to order magnetically near $\sim$ 1 K, were its CF ground state cleanly separated in energy from its excited states \cite{GingrasdenHertog, Gingraslater}.  It displays both Coulomb correlations and a short range ordered spin ice structure at very low temperatures \cite{FennellTb, Fritsch,Fritsch2014, Petit}, as well as novel magnetoelastic modes that couple together acoustic phonons and CF excitations \cite{Fennell2014magnetophonon, Guitteny2013magnetophonon}.  To explain the paradox of Tb$_2$Ti$_2$O$_7$, theory has invoked virtual transitions between the ground state CF doublet and the excited state doublet which stabilize a quantum variant of spin ice, precluding conventional order \cite{Molavian2007PRL, {Bonville2011singlet}}.

The rare-earth stanate pyrochlores, R$_2$Sn$_2$O$_7$, are isostructural to the titanates, and have been of great interest, as they provide a key comparator to the frustrated ground states of the titanates.   Tb$_2$Sn$_2$O$_7$ orders below T$_N \sim 0.88$~K into a ``soft" spin ice ordered phase \cite{MirebeauSN}.  Neutron spectroscopy of Tb$^{3+}$ CF states in Tb$_2$Ti$_2$O$_7$ and  Tb$_2$Sn$_2$O$_7$  show the ground and lowest excited state wavefunctions to both be doublets separated by $\sim$ 1.2 meV in energy.  However the ground state and excited state doublet eigenfunctions are made up of different symmetry allowed m$_J$ values in the two materials, primarily $\sim$ m$_J$=$\pm$5 in the ground state doublet and m$_J$=$\pm$4 in the excited state doublet \cite{Princip_preprint}.  The key point is that the Tb$^{3+}$ eigenfunctions are different in these two sister pyrochlores, even though the low energy eigenvalues are very similar \cite{ZhangCEF, Mirebeau2007}.  As such, solid solutions of Tb$_2$Sn$_{2-x}$Ti$_x$O$_7$ may constitute a novel class of spin liquid, emanating from disorder in the CF wavefunctions, which disrupt the corresponding soft quantum spin ice states.

In this paper we show that strong mixing on the non-magnetic B-site in Tb$_2$B$_2$O$_7$ with B=Ti and Sn yields a novel disordered magnetic ground state.  At variance with the soft quantum spin ice state of the two end members of the solid solution which relies on the proximity of doublet ground and excited states to allow virtual excitations, the novel disordered state at strong B-site mixing originates from a disruption of the CF doublet ground and low lying excited states, such that these states present as extended continua in energy - remarkably, out to three times the bandwidth of these low lying CF excitations in either end member.  It is natural to associate this with a distribution of local B-site environments around each Tb$^{3+}$ ion.  As the CF eigenfunctions generally determine the local ground state moment and anisotropy, the resulting ground state is interpreted as an extreme version of a random anistropy magnet, with random ground state moments, on a pyrochlore lattice.

\begin{figure}
\includegraphics[width=\columnwidth]{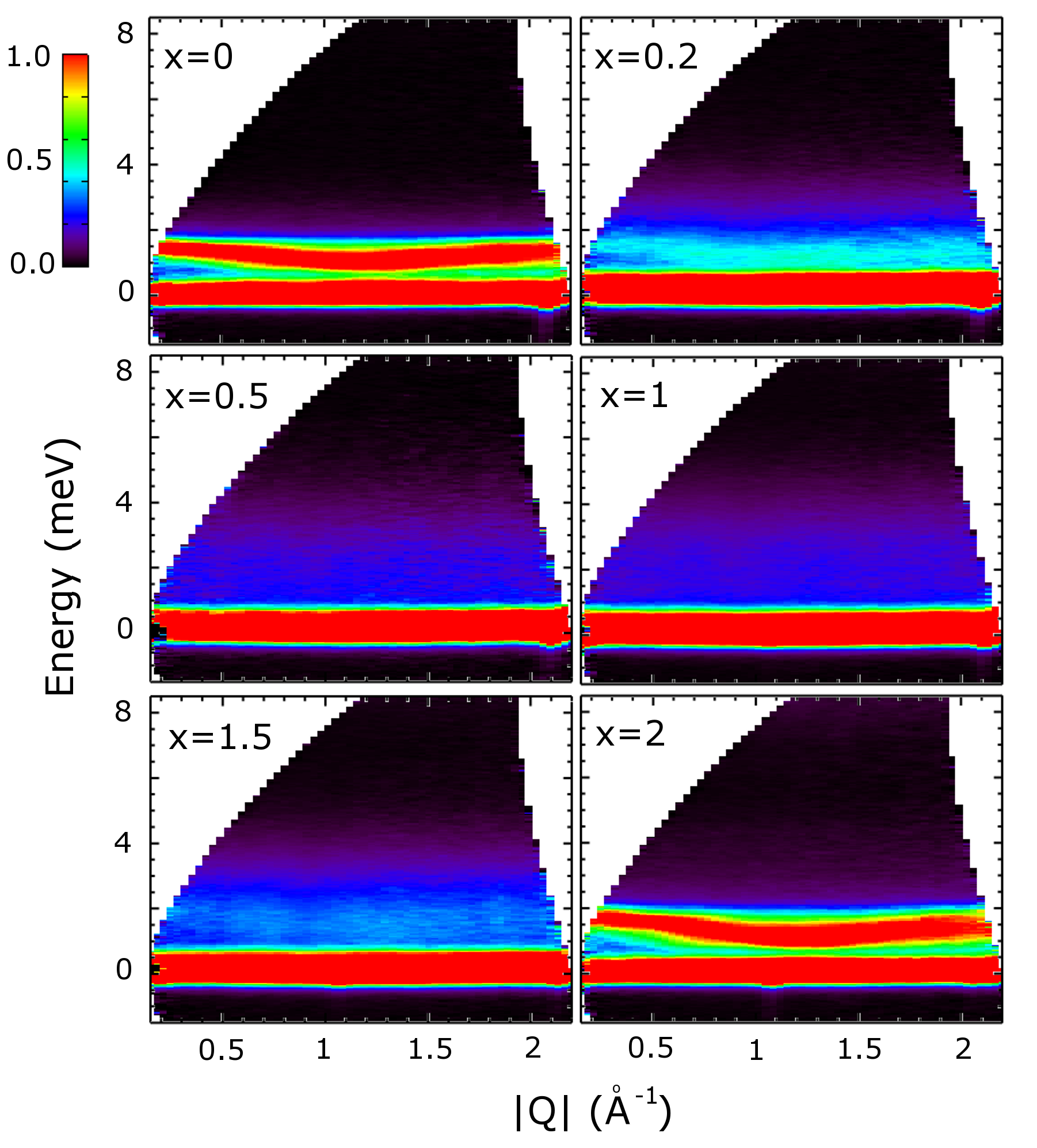}
\caption{\label{fig1} (color online) The measured S($|Q|$,$\hbar\omega$) for Tb$_2$Sn$_{2-x}$Ti$_x$O$_7$ at T=1.5 K is shown.  These low energy, high resolution data was taken with E$_i$=11 meV neutrons.  All data have been corrected for empty can background and detector efficiency.  Well defined transitions between the ground state crystal field doublet and the lowest excited crystal field doublet near 1.5 meV are observed for the end members Tb$_2$Sn$_{2}$O$_7$, $x=0$, and Tb$_2$Ti$_2$O$_7$, $x=2$, while intermediate concentrations see a much broader bandwidth to these low energy excitations. }
\end{figure}

\section{Experimental Details}

Polycrystalline samples of Tb$_2$Sn$_{2-x}$Ti$_x$O$_7$ were prepared by standard solid state synthesis techniques.  These same samples have been studied previously using ac-susceptibility techniques \cite{DahlbergPRB, Matthews_thesis}.  Inelastic neutron scattering measurements were performed on the SEQUOIA direct geometry time-of-flight spectrometer \cite{Granroth2006, Garrett} at the Spallation Neutron Source of Oak Ridge National Laboratory.  Measurements were performed at T=1.5 K and over a wide dynamic range of energies, using incident neutron energies, E$_i$, of 11 meV, 45 meV and 120 meV, in order to probe the appropriate CF excitations and low energy spin dynamics.  

$\mu$SR measurements were performed at the Paul Scherrer Institute using the GPS spectrometer for temperatures greater than 1.6 K and the LTF spectrometer for low temperatures below 1.6 K.  The muon decay asymmetry was recorded as a function of time in all samples in a small longitudinal field (50 Oe) in order to decouple possible low field muon sites.

\section{Neutron Scattering Results}

Figure \ref{fig1} shows colour contour maps of the low energy, inelastic scattering for six different Tb$_2$Sn$_{2-x}$Ti$_x$O$_7$ samples from $x=0$ to $x=2$, using E$_i$=11 meV incident neutrons.  Well defined low energy CF excitations for the end members, $x=0$ and $x=2$, appear as sharp peaks in energy between 1-2 meV with weak dispersion, such that they display a minimum near $|Q|$ $\sim$ 1.25 {\AA}$^{-1}$ for both cases \cite{Gardner99, GardnerPRB, Mirebeau2007, ZhangCEF}.  However, as seen in Fig. \ref{fig1}, the well defined CF excitations are lost at all wavevectors for intermediate concentrations, with some enhanced spectral weight concentrated in the 1-2 meV range for the most lightly disordered $x=0.2$ sample.  Remarkably for $x=0.5$, 1 and 1.5,  we observe a very broad distribution of low energy magnetic scattering at all $|Q|$s studied extending to three times the energy of the lowest lying CF excitations in either end member.

\begin{figure}
\includegraphics[width=0.9\columnwidth]{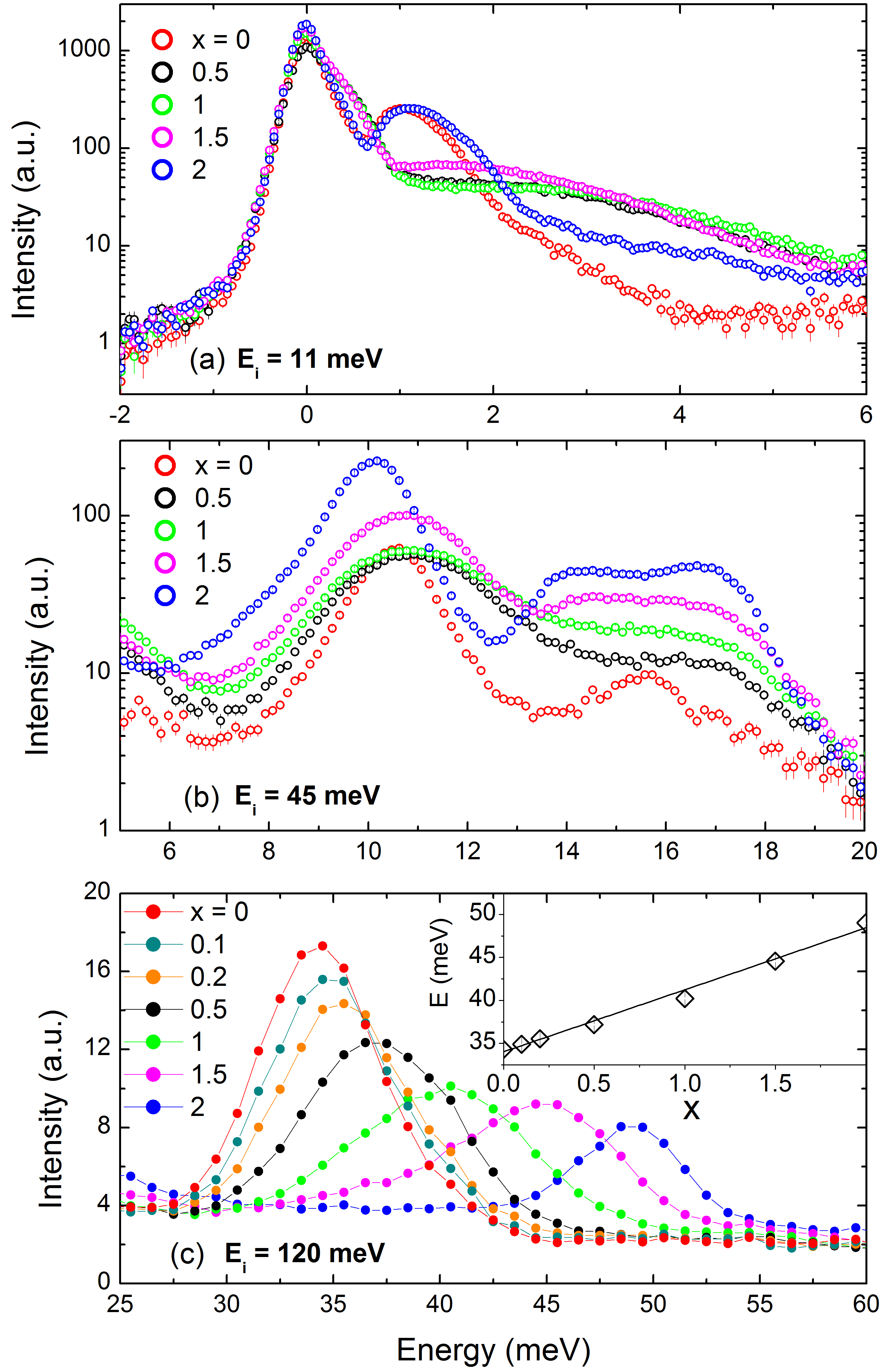}
\caption{\label{fig2} (Color online) Cuts of the inelastic neutron scattering data for Tb$_2$Sn$_{2-x}$Ti$_x$O$_7$ samples at T=1.5 K are shown.  (a) and (b) show inelastic scattering taken with  E$_i$=11 meV and 45 meV neutrons, respectively.  (c) shows inelastic scattering taken with E$_i$=120 meV neutrons.  All data have been corrected for empty can background and detector efficiency.  The high energy crystal field excitations in (c) were fit to extract their energies, and these are plotted in the inset to (c).  Data sets in (a) and (b) have been integrated between 1 {\AA}$^{-1} <|Q|<1.4$ {\AA}$^{-1}$, while that in (c) has been integrated between 2 and 3 {\AA}$^{-1}$ in $|Q|$.}
\end{figure}
Figure \ref{fig2} shows cuts of these and related data sets, taken with different incident energies (E$_i$s) and over different energy ranges, to focus on different CF excitations between 1 meV and 50 meV.  Figure \ref{fig2} (a) shows intensity vs energy cuts for the E$_i$=11 meV data of Fig. \ref{fig1}.  It shows the excitations from the ground state doublet to the excited CF doublets near 1.5 meV in the two end members, as well as the absence of these transitions in the solid solutions.  The $x=0.5$, 1.0, and 1.5 samples all show much enhanced quasi-elastic scattering at low energies ($<$0.8 meV) compared with the end members, and the spectral weight of the diffuse scattering in the regime of the lowest lying CF excitation is now extended to $\sim$ 6 meV, consistent with the colour contour maps of Fig. \ref{fig1}.

At higher energies, i.e. the E$_i$=45 meV data in Fig. \ref{fig2} (b) and E$_i$=120 meV data in Fig. \ref{fig2} (c), we observe a continuous evolution of the CF excitations with concentration.  In the two parent materials the CF excitations in Fig. \ref{fig2} (b) have recently been shown to correspond primarily to transitions from the doublet ground state to singlet CF states at $\sim$ 10, 14 and 17 meV for Tb$_2$Ti$_2$O$_7$, and $\sim$ 10 and 15.5 meV for Tb$_2$Sn$_2$O$_7$, while those in Fig. \ref{fig2} (c) correspond primarily to transitions from the doublet ground state to singlet CF states at $\sim$ 49 meV for Tb$_2$Ti$_2$O$_7$, and $\sim$ 34 and 34.8 meV for Tb$_2$Sn$_2$O$_7$ \cite{ZhangCEF}.  At these higher  energies, where transitions to singlet excited states are primarily studied, we observe a remarkably smooth progression in the B-site disordered systems, as the appropriate CF excitations evolve continuously from the $x=0$ to $x=2$ limit, as seen from $\sim$ 32 meV for Tb$_2$Sn$_{2}$O$_7$ to $\sim$ 49 meV for Tb$_2$Ti$_2$O$_7$ in Fig. \ref{fig2} (c).   The excitations were fit to Lorentzians to extract the CF excitation energy as a function of concentration, and this is shown in the inset to Fig.  \ref{fig2} (c).  Clearly the evolution of this high energy CF excitation is linear with composition.  Furthermore, while the CF excitation broadens somewhat for intermediate compositions, it remains well defined and at energies which are different from that of the end members, very distinct from the transitions between the doublet ground state and doublet first excited states shown in Fig. \ref{fig1} and Fig. \ref{fig2} (a).

\begin{figure}
\includegraphics[width=\columnwidth]{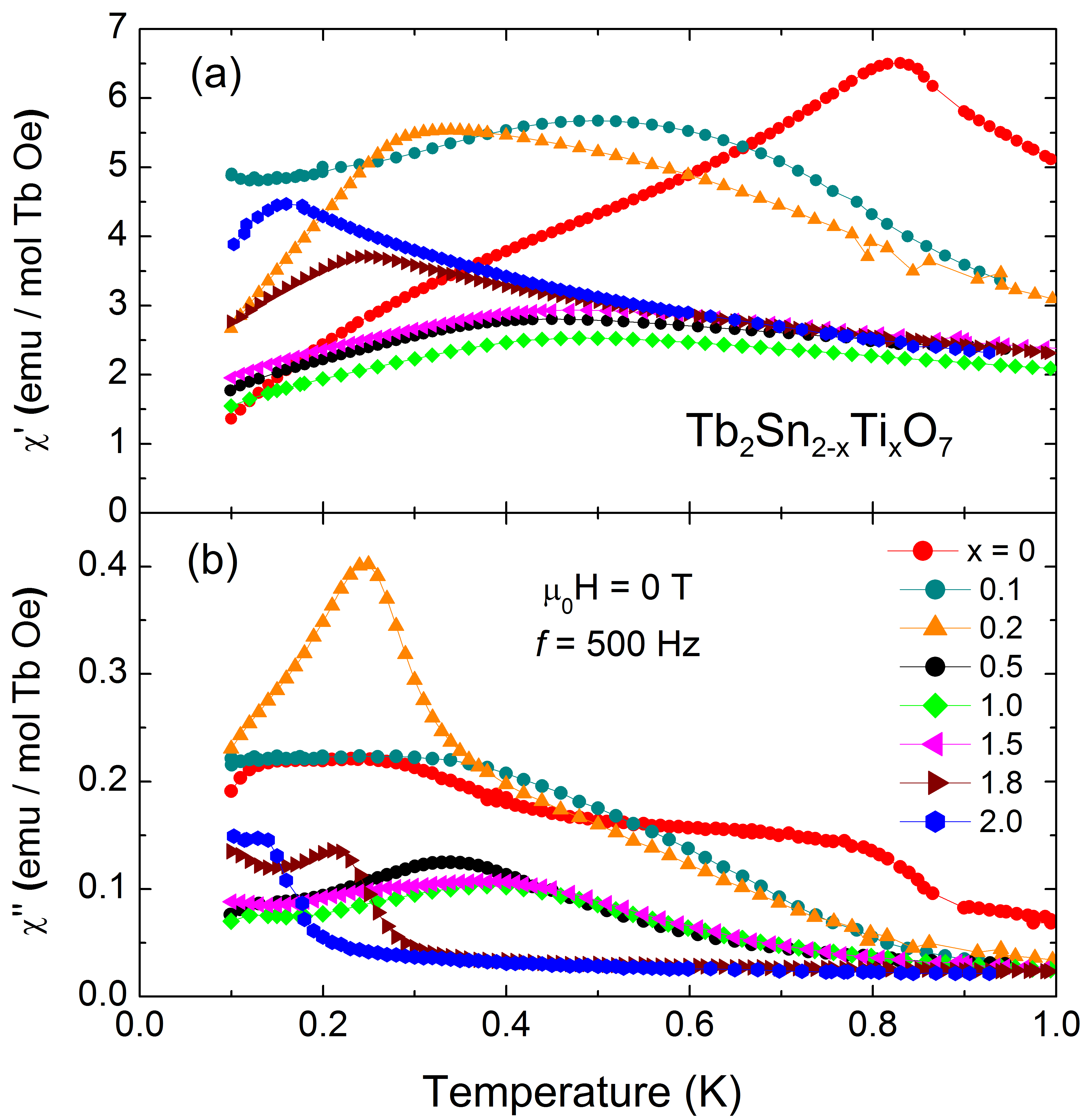}
\caption{\label{fig3} (Color online) The real, $\chi\prime$, and imaginary, $\chi\prime\prime$ parts of the ac susceptibility are shown for Tb$_2$Sn$_{2-x}$Ti$_x$O$_7$ at low temperatures $<$ 1 K.  Measurements were performed in zero DC magnetic field, at a frequency of 500 Hz.}
\end{figure}

\section{AC Susceptibility and $\mu$SR Measurements}

At lower temperatures, the phase behavior and low energy spin dynamics across this series of solid solutions were studied using ac susceptibility and $\mu$SR techniques.    The real, $\chi\prime$, and imaginary, $\chi \prime \prime$, parts of the susceptibility, measured in zero external field and at $f=500$~Hz, are shown in Fig. \ref{fig3} (a) and (b), respectively, for temperatures below 1 K (some of these data and methods have been previously published in Ref.\onlinecite{DahlbergPRB}).  For Tb$_2$Sn$_2$O$_7$ we confirm the peak in $\chi\prime$ and concomitant inflection in $\chi\prime\prime$ at T$_N \sim 0.83$~K, signifying the phase transition to the ``soft'' spin ice ordered phase.  These features are wiped out for all subsequent levels of Sn/Ti mixing, indicating the fragility of the ordered phase in Tb$_2$Sn$_{2}$O$_7$ \cite{DahlbergPRB}.  At the other end of the solid solution, $x=2$, we observe a weak peak in both $\chi\prime$ and $\chi\prime\prime$ indicating glassy behavior for temperatures of $\sim$ 0.23 K and below, consistent with earlier studies \cite{Luo2001,TbTiOglassy1,TbTiOglassy2,TbTiOglassy3,TbTiOglassy4}.  These features display little frequency dependence in contrast to expectations of a canonical spin glass transition \cite{Matthews_thesis}.  However, for strong B-site disorder, in particular $x=0.5$, 1, and 1.5, only very broad features with low absolute values for either $\chi\prime$ or $\chi\prime\prime$, and no tendency of $\chi\prime$  to go to zero as temperature approaches zero, are observed.   This is again consistent with a highly disordered liquid-like ground state distinct from either end member.

\begin{figure}
\includegraphics[width=0.9\columnwidth]{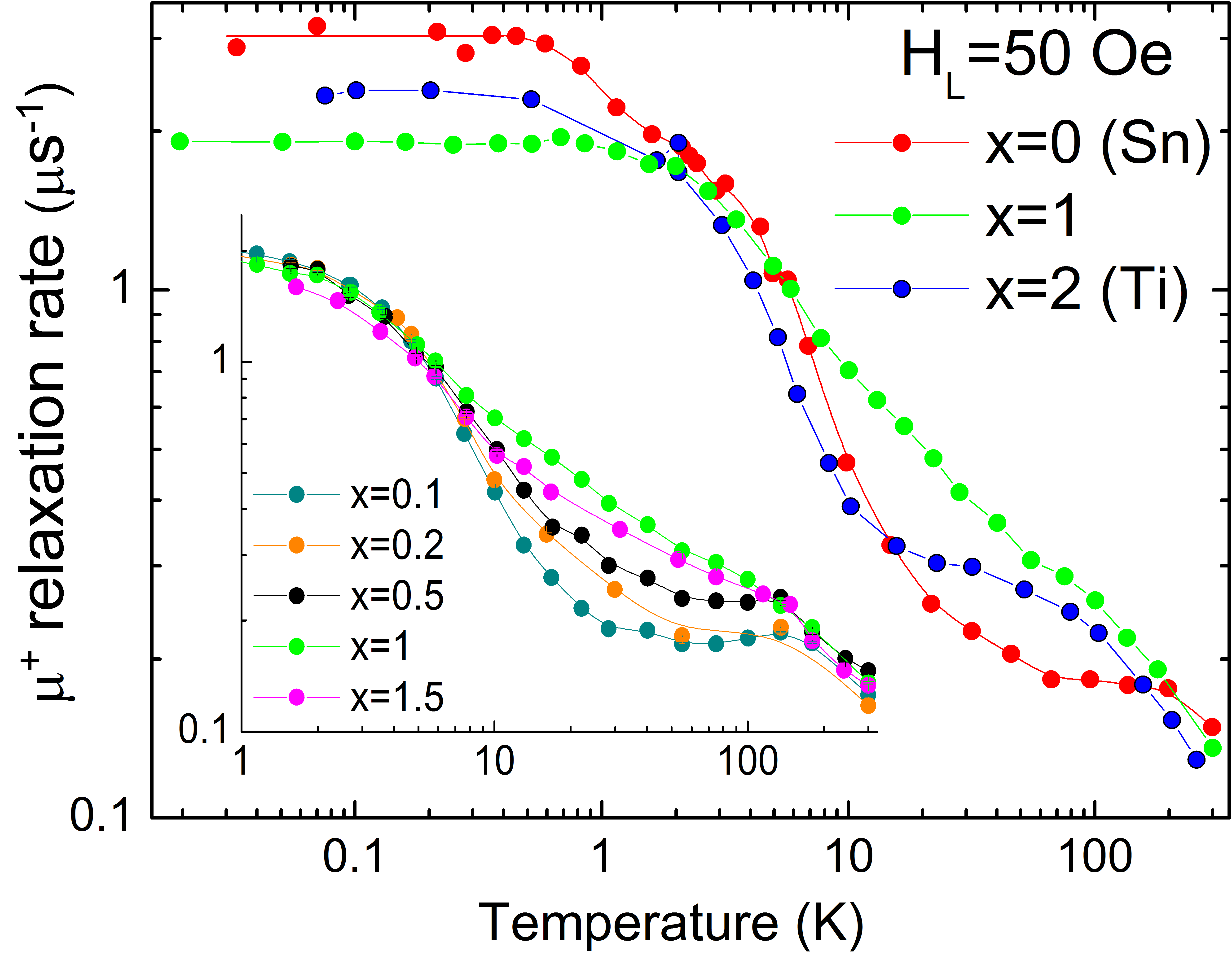}
\caption{\label{fig4} (Color online) The $\mu$SR relaxation rate measured with a small 50 Oe longitudinal field is shown as a function of temperature for selected Tb$_2$Sn$_{2-x}$Ti$_x$O$_7$ samples.  Both the relaxation rate and the temperature are plotted on logarithmic scales.}
\end{figure}

$\mu$SR measurements in a small longitudinal field (50 Oe) were performed on this same solid solution series, spanning the temperature range from 20 mK to 300 K.  We observed a nearly single exponential relaxation to the decay asymmetry over the entire temperature regime and for all samples.  The resulting relaxation times are shown as a function of temperature in Fig. \ref{fig4}.  A weak inflection is observed near the transition to the ``soft'' spin ice ordered phase in Tb$_2$Sn$_{2}$O$_7$, in the temperature dependence to its relaxation rate \cite{deReotier, Bert2006}.   Both Tb$_2$Sn$_{2}$O$_7$ and Tb$_2$Ti$_{2}$O$_7$ display a high temperature plateau between $\sim$ 10 K and 200 K, due to the thermal occupation of the lowest lying CF excited states, $\sim$ 15 K above the ground state for either.  The relaxation rate in Tb$_2$Ti$_{2}$O$_7$ gradually rises to $\sim$ 2.5 $\mu$sec$^{-1}$ at the lowest temperatures measured.  

\begin{figure}
\includegraphics[width=0.9\columnwidth]{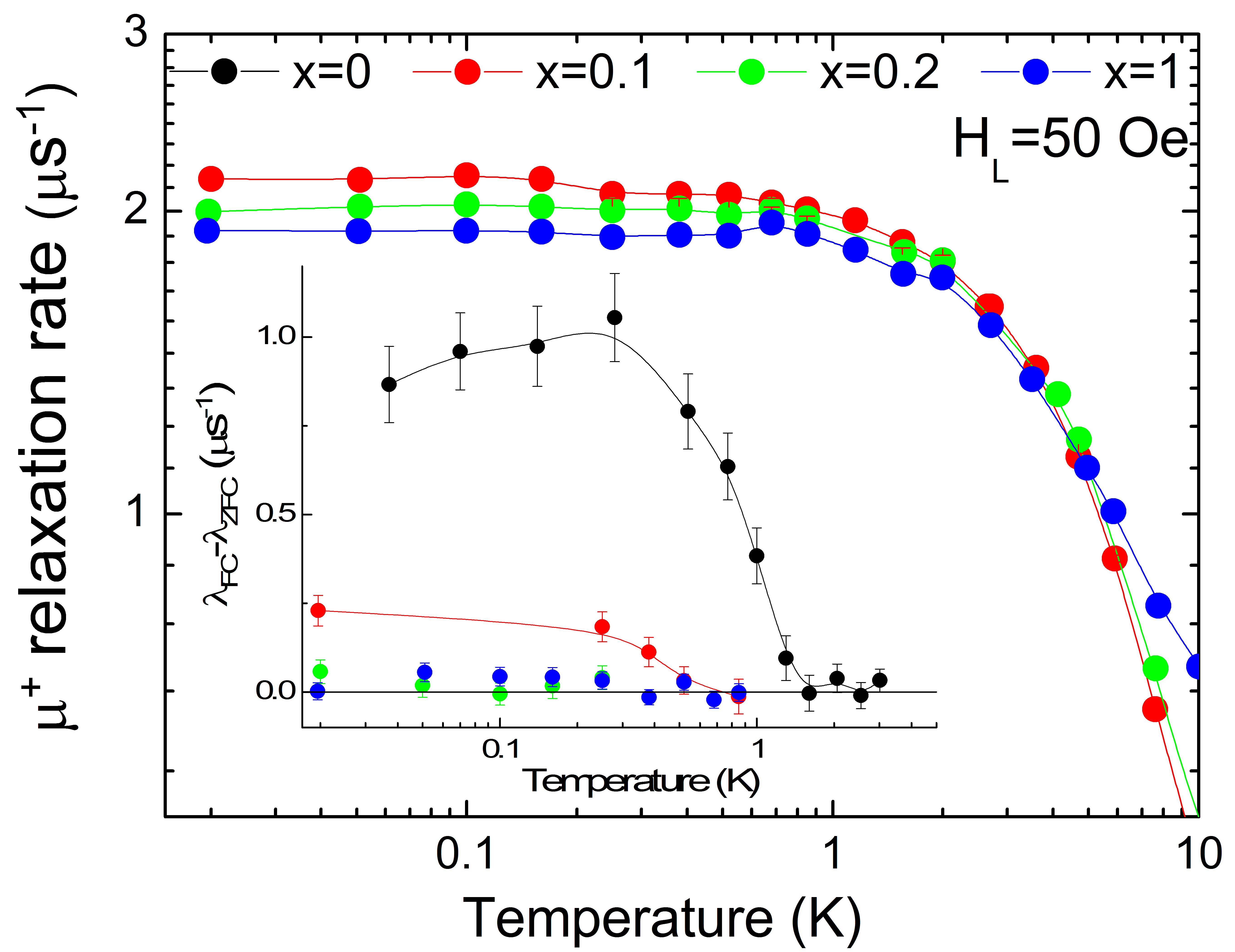}
\caption{\label{fig5} (Color online) The low temperature muon relaxation rates for different members of the Tb$_2$Sn$_{2-x}$Ti$_x$O$_7$ series.  Note that both axes are logarithmic.  The inset shows differences of the relaxation rates measured in a 50 Oe field, on a linear scale, following protocols of a 800 Oe field-cooling (FC) and a zero field cooling (ZFC) for samples with $x=0.1, 0.2$ and 1.0, compared with $x=0$ (from Ref. \onlinecite{Bert2006}).  As with the main panel, the temperature axis is logarithmic. Lines are guides to the eye.}
\end{figure}

The $\mu$SR signature for spin freezing in the ``ordered spin ice" ground state in Tb$_2$Sn$_2$O$_7$ proved to be quite subtle.  None of the usual signs of spin freezing, such as recovery of a 1/3rd tail of the asymmetry or strong sensitivity to an applied longitudinal field, could be detected below T$_C \sim$ 0.9 K \cite{Bert2006, deReotier}.  Indeed the clearest indication for frozen spin correlations was found to be the difference of the relaxation rates measured under field-cooling (FC) or zero field cooling (ZFC) protocols.  We carried out these same FC/ZFC protocols for the Tb$_2$Sn$_{2-x}$Ti$_x$O$_7$ series with $x=0.1, 0.2$ and 1, in addition to $x=0$.  As shown in Fig. \ref{fig5}, this difference, and therefore the ordered spin ice state, is rapidly suppressed at finite $x$ (see inset of Fig. \ref{fig5}).  Both the amplitude of the FC/ZFC difference in relaxation rate, as well as its onset temperature, are reduced from 1.2 K at $x=0$ to $\sim$ 0.5 K for $x=0.1$, a temperature which corresponds to a broad maximum in the ac susceptibility (Fig. \ref{fig3} (a)).  This FC/ZFC signature completely disappears for $x=0.2$ and $x=1.0$.  While $\chi '$ displays a peak as a function of temperature at $\sim$ 0.25 K, possibly indicative of a spin glass transition, only weak features are observed for $0.2 < x < 1.8$.  We therefore conclude that the ground state for Tb$_2$Sn$_{2-x}$Ti$_x$O$_7$ and $0.2 < x < 1.8$ is a fully dynamical spin liquid.

\begin{figure}
\includegraphics[width=0.9\columnwidth]{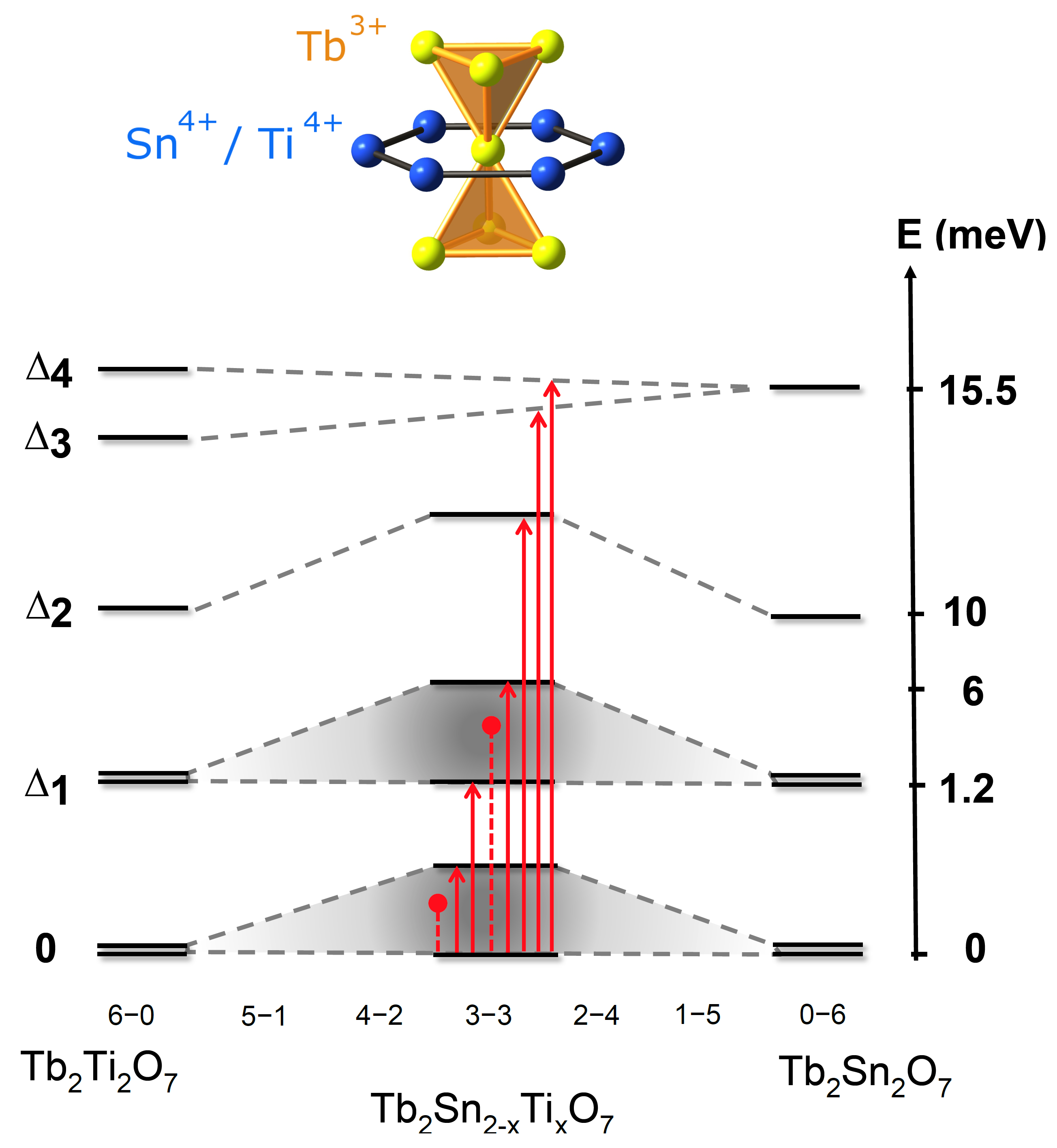}
\caption{\label{fig6} (Color online) A minimal model for the effect of B-site disorder on the low energy CF levels in Tb$_2$B$_2$O$_7$ is illustrated.  B-site disorder can split the doublet CF states, as Tb$^{3+}$ is a non-Kramers ion, but the singlets are simply shifted in energy.  Disordered B-site samples possess a distribution of the seven variants (3 Sn$^{4+}$, 3 Ti$^{4+}$ ions, 4 Sn$^{4+}$, 2 Ti$^{4+}$ ions, etc.) of local CF environment at the Tb$^{3+}$ site. Transitions from the ground state to excited CF states are indicated with red arrows, while those terminating with a red dot indicate inelastic peak positions expected for high B-site disorder, due to averaging over local Tb$^{3+}$ environments. The local environment at the Tb$^{3+}$ site in the pyrochlore structure is indicated at the top.  For clarity, the six O$^{2-}$ ions around each Tb$^{3+}$ ion are omitted.}
\end{figure}
For intermediate concentrations, the high temperature plateau fills in (see inset to Fig. \ref{fig4}), consistent with the effect of strong B-site disorder on the low lying crystal field levels, shown in Figs \ref{fig1} and \ref{fig2}.   Again at low temperatures the relaxation rate rises gradually to a low temperature limit of $\sim$ 2 $\mu$sec$^{-1}$, showing a progressively smaller low temperature limit than either end member; again consistent with a disordered and dynamic ground state, and no hysteretic behavior, in the presence of strong B-site disorder.

\section{Discussion and Conclusions}

The Tb$^{3+}$ ion possesses an even number of electrons, in contrast to Dy$^{3+}$ and Yb$^{3+}$, and thus its CF configuration is not protected by Kramers theorem.  The 13 CF levels in Tb$_2$Ti$_{2}$O$_7$ and Tb$_2$Sn$_{2}$O$_7$ then appear as both singlets and doublets \cite{ZhangCEF}.  Remarkably, our neutron spectroscopy of the CF states in the presence of B-site disorder strongly suggests that doublet states present as extended continua in energy, while singlet states have their energy eigenvalues and lifetimes modified, but appear less strongly affected.   This is consistent with a minimal model of the low lying CF levels, where CF doublets split in the presence of B-site disorder, but singlets simply have their eigenvalues modified, as illustrated in Fig. \ref{fig6}.  As the local crystalline environment around a Tb$^{3+}$ ion, shown at the top of Fig. \ref{fig6}, is determined by a ring of B-site ions and associated O$^{2-}$ ions, a strongly disordered  Tb$_2$Sn$_{2-x}$Ti$_x$O$_7$ sample would display the appropriate statistical distribution of environments for the Tb$^{3+}$ sites, corresponding to all permutations of Ti$^{4+}$ and Sn$^{4+}$ ions decorating the six neighboring B-sites.  The B-site ions have different ionic radii, and would also perturb the local O$^{2-}$ positions slightly.  The CF eigenfunctions and eigenvalues determine both the local Tb$^{3+}$ ground state moment and its anisotropy.  Such a distribution of quenched, disordered CF environments would give an extreme version of a random anisotropy magnet, in the presence of random moment sizes, which, at a minimum, would also disorder the dipolar part of the spin Hamiltonian in Tb$_2$Sn$_{2-x}$Ti$_x$O$_7$.  A related effect on the doublet CF ground state of Pr$^{3+}$ in A-site disordered Pr$_{2-x}$Bi$_x$Ru$_2$O$_7$ has likely been observed \cite{vanDuijn}.

Taken together, these experimental results show a remarkable route to a new type of strongly fluctuating magnetic liquid state, based on strong quenched disorder of the Tb$^{3+}$ CF environment.  Given that both the Sn$^{4+}$ and Ti$^{4+}$ sites are non-magnetic, one may naively expect the mixing of these elements on the B-sites of Tb$_2$B$_2$O$_7$ to have little effect on the low temperature physics displayed across the solid solution.  Yet it has a dramatic effect; it wipes out the subtle transition to a ``soft'' spin ice ordered phase in Tb$_2$Sn$_{2}$O$_7$, and produces a fluctuating spin liquid state at intermediate concentrations.

These results also underline the crucial role played by the low lying excited CF doublet for Tb$^{3+}$ in both end members, $x=0$ and $x=2$.  The modification of the ground state and lowest excited state doublet wavefunctions on moving from $x=0$ to $x=2$, implies a disruptive change to the virtual excitations between the ground state and low lying excited states.  As this is the mechanism proposed \cite{Gingraslater} by which quantum fluctuations are introduced to  Tb$_2$Ti$_{2}$O$_7$, and which presumably are also at play in Tb$_2$Sn$_{2}$O$_7$, the advanced characterization of the Tb$_2$Sn$_{2-x}$Ti$_x$O$_7$ solid solutions provides an important perspective on such quantum fluctuations in the presence of relevant disorder. 

%\begin{scilastnote}
%\item[]
We acknowledge helpful discussions with Michel Gingras.  Work at McMaster University was supported by NSERC of Canada.  Materials synthesis was supported by the US DOE grant DG-FG02-08ER46544.  M.L.D., M.J.M., and P.S. would like to acknowledge support from NSF grant No. DMR-070158, No. DMR-1104122 and No. DMR-1341793. This work was partly supported by grants ANR-09-JCJC-0093-01, ANR-SPINLIQ-86998 and by the EU FP-7 under the NMI3-II grant number 283883.  Research conducted at ORNL's Spallation Neutron Source was sponsored by the Scientific User Facilities Division, Office of Basic Energy Sciences, US Department of Energy.
%\end{scilastnote}

%\bibliography{pyro_mar12}

\end{document}